\documentclass[10pt]{book}
\usepackage{array}
\usepackage{amsthm,amsmath,amssymb}
\usepackage{natbib}
\usepackage{graphicx}
\usepackage{multirow}

\newtheorem{theorem}{\bf{Theorem}}
\newtheorem{remark}{\bf{Remark}}
\newtheorem{lemma}{\bf{Lemma}}

\newtheorem{corol}{\bf{Corollary}}

\newtheorem{proposition}{\bf{Proposition}}

\begin{document}
\renewcommand{\baselinestretch}{1.2}
\markright{
}
\markboth{\hfill{\footnotesize\rm DAVID DEGRAS
}\hfill}
{\hfill {\footnotesize\rm ROTATION SAMPLING FOR FUNCTIONAL DATA} \hfill}
\renewcommand{\thefootnote}{}
$\ $\par
\fontsize{10.95}{14pt plus.8pt minus .6pt}\selectfont
\vspace{0.8pc}
\centerline{\large\bf ROTATION SAMPLING FOR FUNCTIONAL DATA}
\vspace{.4cm}
\centerline{David Degras}
\vspace{.4cm}
\centerline{\it DePaul University}
\vspace{.55cm}
\fontsize{9}{11.5pt plus.8pt minus .6pt}\selectfont

\begin{quotation}
\noindent {\it Abstract:}
This paper addresses the survey estimation of a population mean in continuous time. For this purpose we extend the rotation sampling method to functional data. In contrast to conventional rotation designs that select the sample before the survey, our approach randomizes each sample replacement and thus allows for adaptive sampling. Using Markov chain theory, we evaluate the covariance structure and the integrated squared error [ISE] of the related Horvitz-Thompson estimator. Our sampling designs decrease the mean ISE by suitably reallocating the sample across population strata during replacements. They also reduce the variance of the ISE by increasing the frequency or the intensity of replacements. To investigate the benefits of using both current and past measurements in the estimation, we develop a new composite estimator. In an application to electricity usage  data, our rotation method outperforms fixed panels and conventional rotation samples. Because of the weak temporal dependence of the data, the composite estimator only slightly improves upon the Horvitz-Thompson estimator. 
\par

\vspace{9pt}
\par
\end{quotation}\par


\fontsize{10.95}{14pt plus.8pt minus .6pt}\selectfont
\setcounter{chapter}{1}
\setcounter{equation}{0} 
\noindent {\bf 1. Introduction}

\smallskip

In various monitoring applications, sensor networks generate large volumes of data in continuous time. Due to cost or energy constraints, these collections of functional data (that is, curve data) often cannot be exhaustively observed. Electric utilities, for instance, need to monitor their clientsÕ total consumption in order to adjust the power generation to the system load, to predict future consumption, and to determine pricing policies. However, they cannot access all clientsÕ smart meters at each instant since this would exceed the network transmission capacity and/or incur considerable costs. Under such observation constraints, survey sampling provides competitive solutions for monitoring global parameters. (See Chiky, CubillŽ, Dessertaine, H\'ebrail, and Picard (2008) for a comparison between survey sampling and signal compression approaches).

In various monitoring applications, sensor networks generate large volumes of data in continuous time. Due to cost or energy constraints, these collections of functional data (that is, curve data) often cannot be exhaustively observed. Electric utilities, for instance, need to monitor their clientsÕ total consumption in order to adjust the power generation to the system load, to predict future consumption, and to determine pricing policies. However, they cannot access all clientsÕ smart meters at each instant since this would exceed the network transmission capacity and/or incur considerable costs. Under such observation constraints, survey sampling provides competitive solutions for monitoring global parameters. (See Chiky, Cubill\'e, Dessertaine, H\'ebrail, and Picard (2008) for a comparison between survey sampling and signal compression approaches.) 

Several recent studies explore survey estimation based on functional data. Cardot, Degras, and Josserand (2012) extend the Horvitz-Thompson [HT] estimator to functional data and construct simultaneous confidence bands for the population mean function based on results of Degras (2011). Cardot, Chaouch, Goga, and Labru\`ere (2010) investigate functional principal component analysis in design-based surveys and apply this technique to integrate auxiliary information. Cardot, Goga and Lardin (2013) develop model-assisted survey estimators for functional data. All of these studies rely on fixed panel designs: the same sample is used throughout the survey. Rotation sampling, in contrast, typically yields more accurate estimates of population parameters. This method, which replaces part of the sample at each survey occasion, is widely used in practice (e.g., U.S. Current Population Survey) and has received considerable attention in the literature (e.g., Eckler, 1955; Rao and Graham, 1964; Wolter, 1979; and Lavall\'ee, 1995). However, the available studies rely on modeling frameworks (e.g., discrete time, infinite population and stationary measurements) that are unsuitable for continuous-time monitoring applications and functional data. 

In this paper we investigate rotation sampling for functional data. First, we devise sampling designs that replace the sample in part or in full at prespecified times. In contrast to conventional rotation designs, which determine the sample prior to the survey, our approach randomizes each sample replacement. The sample can thus be adaptively selected, i.e. it can be improved over time based on the observed data. (See Thompson and Seber, 1996, for a comparison between adaptive sampling designs and conventional designs.) Second, we study the HT estimator of the mean function in a stratified population. Using Markov chain theory, we derive large-sample approximations for the mean and variance of the integrated squared error [ISE]. If sample sizes in each stratum are constant over time, rotation samples and fixed panels have on average the same ISE. However, our rotation designs can reduce the mean ISE by suitably reallocating the sample at each replacement time (Neyman allocation). In addition, rotation samples dramatically decrease the variance of the ISE in comparison to fixed panels. Third, we develop a composite estimation procedure (see e.g., Rao and Graham, 1964) in order to improve upon the HT estimator. The composite estimator is recursively defined in terms of its value at an arbitrary previous instant, the estimated change in the population mean, and the HT estimator. Finally, we apply our sampling strategies to electricity usage data from the Irish CER Smart Metering Project (CER, 2011). The numerical study confirms that our rotation designs outperform fixed panels and conventional rotation samples both in terms of estimation accuracy and stability. The composite estimator slightly improves upon the HT estimator. 

The paper is organized as follows. In Section 2, we present the modeling framework and the HT estimator. In Section 3, we define the new rotation sampling designs. The mean and covariance of the HT estimator are studied in Section 4. Section 5 gives the main results on the variance of the ISE. Section 6 introduces the composite estimator. The numerical study is described in Section 7. Section 8 provides concluding remarks. The main proofs are gathered in the Appendix. Additional proofs are available online as supplementary material.

\bigskip


\setcounter{chapter}{2}
\setcounter{equation}{0} 
\noindent {\bf 2.  Statistical framework}
\smallskip

Let $U_N=\{ 1,\ldots, N \}$ be a finite population and let   $X_k\, , k\in U_N , $
 be deterministic functions defined on a bounded interval $[0,T]$.   
We study the estimation of the population mean function
\[ \mu_N(t) = \frac{1}{N} \sum_{k\in U_N} X_k(t)   \] 
based on $X_k(t), \, k\in s(t)$, where $ t\in[0,T]$ and 
$s(t) \subset U_N$ is a probability sample of fixed size $n(t)$. 
The collection of samples $s=\{ s(t): t\in[0,T] \}$ 
can be viewed as a random function from  $[0,T]$ to  the set $\mathcal{P}(U_N)$ of all subsets of $U_N$. 
It is selected from 
the function space $\mathcal{S}=\{ s: [0,T] \to \mathcal{P}(U_N) \}$ 
according to a probability measure $P$ specified by the statistician. 
For all $k,l\in U_N$ and $ t ,t' \in[ 0,T]$, we denote 
the first and second order inclusion probabilities by 
$\pi_k(t)= P(k \in s(t)) = P( \{ s \in \mathcal{S}: k \in s(t)\})$  
and $\pi_{kl} (t,t')= P(k\in s(t),\, l \in s(t')) \break =  P(\{s \in \mathcal{S}: k \in s(t), l\in s(t')\})$. 
Throughout the paper all expectations 
are taken with respect to $P$. 

 We consider the  estimator of  Horvitz and Thompson (1952)  
 \begin{equation}\label{HT estimator}
\hat{\mu}_{ht} (t) = \frac{1}{N} \sum_{k \in U_N } \frac{I_k(t)}{\pi_k (t)}\, X_k(t)  ,
\end{equation}
where  
 $I_k (t)$ is the sample indicator function of $k\in U_N$ at time $t$: 
 $I_k(t)=1$ if $k\in s(t)$ and $I_k(t)=0$ otherwise.  
This estimator, which we refer to as the HT estimator, 
is unbiased for $\mu_N(t)$ 
and its 
 covariance function is 
\[
 \mathrm{Cov}(\hat\mu_{ht}(t),\hat\mu_{ht}(t')) = \frac{1}{N^2} \sum_{k,l\in U} \frac{\Delta_{kl}(t,t')}{\pi_k(t)\pi_l(t')} X_k(t)X_l(t')  ,
\]
where $\Delta_{kl}(t,t') = \mathrm{Cov}(I_k(t),I_l(t')) = \pi_{kl}(t,t') -\pi_k(t) \pi_l(t')$. 

The estimation accuracy can often be improved by stratifying the population. 
From now on   we drop the subscript in  $U_N$ and 
assume that $U$ is partitioned into strata $U_{h} ,\,  1\le h\le H,$ of size $N_h$. 
We denote  the strata mean  and covariance functions
 by  
 $ \mu_{h}(t) = (1/N_h) \sum_{k\in U_{h}} X_k(t) $ and 
 \[ \gamma_{h}(t,t') = \frac{1}{N_h-1} \sum_{k\in U_{h}} \left( X_k ( t) - \mu_{h}(t) \right)  \left( X_k ( t') - \mu_{h}(t') \right). \] 
Let $n_h(t) = \#(s(t)\cap U_h)$ be the sample size in $U_h$ at time $t$ and 
$f_{h}(t)=n_h(t)/N_h$ be the sampling rate. 
If $s(t)$ is obtained by simple random sampling without replacement [SRSWOR] 
independently in each $U_h$, 
the HT estimator becomes  
\begin{equation}\label{HT strat time-varying}
\hat{\mu}_{ht} (t)= \frac{1}{N} \sum_{h=1}^H  \frac{1}{f_h(t)} \sum_{k \in U_h } I_k (t) X_k(t) 
\end{equation}
 and its covariance rewrites as
\begin{equation}\label{covariance time-varying strat}
\mathrm{Cov}\big(\hat{\mu}_{ht}(t),\hat{\mu}_{ht}(t')\big) = 
\frac{1}{N^2}  \sum_{h=1}^H  \frac{1}{f_h(t)f_h(t')} \sum_{k,l\in U_h}  \Delta_{kl}(t,t') X_k(t) X_l(t')  .
\end{equation}

\begin{remark}
For simplicity, we assume that the sampled curves $X_k$ 
are observed in continuous time and without noise. 
If  these curves are observed at discrete times and/or with noise,
 interpolation or smoothing methods should be applied. 
 In this case the results of this paper still hold under  standard interpolation 
 or smoothing conditions.
 See for example Cardot and Josserand (2011) and Cardot, Degras and Josserand (2012).  

\end{remark}

\medskip


\setcounter{chapter}{3}
\setcounter{equation}{0} 
\noindent {\bf 3. Rotation designs for continuous-time surveys}

\smallskip

Rotation sampling has so far been developed for discrete-time surveys. 
In this section we extend it  to the continuous-time framework of functional data. 
We propose two sampling designs (i.e., two choices of the probability measure $P$)
for selecting the time-varying sample $s=\{ s(t):  t\in[0,T] \}$ in a stratified population. 
These sampling designs, which we refer to as full replacement and partial replacement, share the following features: 

\begin{itemize}

\item The time-varying samples $s_h =\{ s(t)\cap U_h:  t\in[0,T]\},  \, 1\le h \le H, $  are independent across strata. 

\item At time $\tau_0=0$, the samples $s_h(\tau_0)$ are obtained by SRSWOR. 

\item The $s_h $ can be modified at fixed times  $0  < \tau_1 < \ldots < \tau_m < T$. 

\end{itemize}

\noindent It remains to specify the probability distribution of the discrete processes $\{ s_h(\tau_r): 1\le r \le m \} $ 
under full and partial replacement.

\begin{enumerate}
\item {\bf  Full replacement.} 
For each $h$, the successive samples $s_h(\tau_r), 1\le r \le m,$ are obtained  by independent SRSWOR 
 of $n_h(\tau_r) $ units  in $U_h$.

\item {\bf Partial replacement.} 
For each $h$, a fraction $\alpha_h \in [0,1]$ of $s_h(t)$ is replaced at each time $\tau_r, \, 1\le r \le m$. 
More precisely,  given $s_h(\tau_{r-1})$, $s_h(\tau_{r})$ is obtained by the following independent operations:   
\begin{itemize}
\item 
select  $\alpha_h n_h(\tau_{r-1})$ units   in $s_h(\tau_{r-1})$ by SRSWOR and discard them from the sample;

\item select $(n_h(\tau_r) - (1-\alpha_h) n_h(\tau_{r-1})  )$ units  in $U_h \setminus s_h(\tau_{r-1})$ by SRSWOR 
and add them to the sample.
\end{itemize}

\end{enumerate}
By construction, the process $\{ s_h(\tau_r):0\le r \le m \} $ is a Markov chain both under   full and partial replacement. 
Note that full replacement is not a special case of partial replacement with $\alpha_h=1$: indeed, 
$s_h(\tau_{r-1})$ and $s_h(\tau_r)$ are independent in the former case whereas they are disjoint (and thus dependent) in the latter. 
In partial replacement we refer to the $ \alpha_h $ as the replacement rates. For simplicity we assume that the $\alpha_h$ are constant over time and that the proposed sample replacements are possible without modifications, which entails that $\alpha_h n_h(\tau_{r-1}) \in \mathbb{N}$ and $n_h(\tau_{r-1}) \le  n_h(\tau_r)+    \alpha_h n_h(\tau_{r-1})  \le N_h $ for all $h,r$. Fixed panels correspond to partial replacement with $\alpha_h=0$ and 
the $n_h(t)$ constant over time.

We now determine the probability distribution of the sample $s_h(t)$ under the proposed designs.
The following result relies on an induction argument on the $\tau_r$ under partial replacement;  
it holds trivially under full replacement.
\begin{proposition}\label{time-varying inclusion probabilities}
Assume either the full or the partial replacement design.  
For all $1\le h \le H$ and $ t\in[0,T]$, 
the probability distribution of $s_h(t)$ is identical to the SRSWOR of $n_h(t)$ units in $U_h$. 
\end{proposition}

\bigskip


\setcounter{chapter}{4}
\setcounter{equation}{0} 
\noindent {\bf 4. Covariance of the Horvitz-Thompson estimator}
\smallskip

Here we derive the covariance function \eqref{covariance time-varying strat} of the HT estimator (\ref{HT strat time-varying}) under 
the previous rotation designs, which amounts to determining  $\Delta_{kl}(t,t')$ explicitly. 
%
Let  $  \nu(t) = \min\left\{ r:   \tau_r \le t \right\} $
be the number of sample replacements before time $t$.  
For $0 \le t < T$, it holds that $ \tau_{\nu(t)} \le t < \tau_{\nu(t)+1} $. 
By convention, we set $\tau_{m+1}=T$ and $\nu(T)=m$. 
Let $ \delta_{\cdot\cdot}$ indicate the  Kr\"onecker delta. 


\medskip

\noindent {\bf 4.1. Covariance under full replacement}

\smallskip

Under the full replacement design, $\hat{\mu}_{ht}(t)$ and $\hat{\mu}_{ht}(t')$  are independent if the sample has been replaced between times $t$ and $t'$. 
If no replacement occurred between $t$ and $t'$,  $\Delta_{kl}(t,t')$ 
can be derived from the properties of SRSWOR. 

 \begin{theorem}\label{covariance estimator: full}
Assume the full replacement design.  
For all  $    t ,  t' \in[0,  T]$, 
  \[
\mathrm{Cov}\big(\hat{\mu}_{ht}(t),\hat{\mu}_{ht}(t')\big) = 
 \frac{1}{N}  \sum_{h=1}^H  \frac{N_h}{N}  \, \frac{1-f_{h}(t)}{f_{h}(t)}\, \gamma_{h}(t,t') \,  \delta_{\nu(t)\nu(t')}   \, .
\]
\end{theorem}

This theorem will be commented in relation to partial replacement in the next section.

\medskip
\pagebreak


\noindent{\bf 4.2. Covariance under partial replacement}
\label{sec: cov part rep}

\smallskip

To derive $\Delta_{kl}(t,t')$, 
it suffices to find  $\pi_{kl}(t,t')$ in view of Proposition \ref{time-varying inclusion probabilities}.   
By definition of SRSWOR,  for a given stratum $U_h$, 
$\pi_{kk}(t, t')  $ and $\pi_{kl}(t, t')  $  
do not depend on $k, l \in U_h$ ($k\ne l$). 
Since $\sum_{k\in U_h}I_k(t) = n_h(t)$, it follows that
$E(\sum_{k}I_k(t) \sum_{l}I_l(t') )= n_h(t)n_h(t') = N_h  \pi_{kk}(t,t') + N_h(N_h-1) \pi_{kl}(t,t') $, 
with 
$k \ne l $ two arbitrary units  in $U_h$. 
Therefore, it suffices to determine  $\pi_{k k}(t, t') $.  
This in turn reduces to computing $P  ( k \in s_h(t')  | k \in s_h(t))$.  

Let  $D$ be a subset of  $U_h$.  
The Markovian nature of $\{ s_h(\tau_r): 0\le r \le  m\}$ and the properties of SRSWOR 
(namely, the probability that the sample contains $D$ only depends on the size of $D$) yield the following result. 

\begin{lemma}\label{lem: s inter D MC}
Under the partial replacement design,  
$\{ s_h(\tau_r)\cap D : 0\le r \le m\}$ is a Markov chain. 
\end{lemma}

By setting $D=\{k\}$ 
in Lemma \ref{lem: s inter D MC}, it stems that 
 $\{ I_k(\tau_r) : 0 \le r \le  m \}$ is a Markov chain whose    
transition probabilities 
can be found with the Chapman-Kolmogorov equations. 
To this end, define 
\begin{equation}\label{lambda}
\lambda_{h}(t,t') = 
 \prod_{r=\nu(t)+1}^{\nu(t')} \frac{1-\alpha_h - f_{h}(\tau_r)}{1-f_{h}(\tau_{r-1})} 
\end{equation} 
for $0\le  t \le t' \le T$ with $\lambda_h(t,t')=1$ if $\nu(t)=\nu(t')$. 
Set to 1 all factors in $\lambda_h(t,t')$ for which $f_{h}(\tau_{r-1})=1$ 
and extend $\lambda_h(t,t')$ as a symmetric function on $[0,T]^2$. 
 
%


\begin{lemma}\label{r-step transition probability}
Assume the partial replacement design.  
For all $U_h$, $k\in U_h$, and $0\le  t \le t' \le T$, 
\[
\left\{ 
\begin{array}{l}
P \left( k \in s_h(t') \big| k\in s_h(t) \right) = \left(1-f_{h}(t)\right) \lambda_h(t,t')+ f_{h}(t') ,\\
\displaystyle P \left( k \in s_h(t') \big| k\notin s_h(t) \right) = f_{h}(t') - f_{h}(t)\lambda_h(t,t').
\end{array}
\right. 
\]
\end{lemma}
The lemma is easily proved by induction and, 
with simple matrix diagonalizations, $\lambda_h(t,t') $ expresses as the product of the eigenvalues 
of the transition probability matrices of  $\{ I_k(\tau_r) : 0 \le r \le  m \}$ between $t$ and $t'$. 

For any two real numbers $x,y,$ write $x\wedge y = \min(x,y)$ and $x \vee y = \max(x,y)$. 
Based on Proposition \ref{time-varying inclusion probabilities} and Lemma \ref{r-step transition probability}, 
we obtain the covariance function  (\ref{covariance time-varying strat}). 

 \begin{theorem}\label{covariance estimator: partial}
Assume the partial replacement design.
For all 
$    t,  t' \in[0, T]$, 
\[
\mathrm{Cov}\big(\hat{\mu}_{ht}(t),\hat{\mu}_{ht}(t')\big) = 
 \frac{1}{N}  \sum_{h=1}^H  \frac{N_h}{N} \,\frac{1  - f_{h}(t\wedge t'  )}{f_{h}(t\vee t')} \, \gamma_{h}(t,t') \, \lambda_h(t,t')   .
\]

\end{theorem}

\medskip

To gain insight into Theorems \ref{covariance estimator: full}-\ref{covariance estimator: partial}, we suppose that the  $n_h(t)$ are constant over time. 
In the case of fixed panels (partial replacement with $\alpha_h=0$), 
$\mathrm{Cov}(\hat{\mu}_{ht}(t),\hat{\mu}_{ht}(t')) = N^{-2}\sum_h N_h (f_h^{-1}-1) \gamma_h(t,t')$ for all $t,t'$. 
Under full replacement, 
the estimator covariance is the same as for fixed panels on the diagonal blocks $ [\tau_r ,\tau_{r+1}]^2 ,1\le r \le m,$ and is zero outside these blocks. 
Under partial replacement, the term $\lambda_h(t,t')$ simplifies to $  \left( 1 - \alpha_h/(1 - f_h) \right)^{|\nu(t)-\nu(t')|} $. 
Hence for fixed times $t,t'$, the correlation between $\hat{\mu}_{ht}(t)$ and $\hat{\mu}_{ht}(t')$ decreases as 
 $\alpha_h\in  [0,1-f_h]$ increases (assuming $\gamma_h(t,t')>0$).  
If $\alpha_h=1-f_h$ for all $h$, the covariance is the same as under full replacement. 
For  values $\alpha_h > 1- f_h$, 
the covariance becomes unstable in the sense that  $\lambda_h(t,t')$
 changes sign on every block $ [\tau_q ,\tau_{q+1}]\times  [\tau_r ,\tau_{r+1}]$. 
If $\alpha_h\notin\{0, 1-f_h\}$ for all $h$, 
$|\mathrm{Cov}(\hat{\mu}_{ht}(t),\hat{\mu}_{ht}(t'))|$ decreases at an exponential rate as $|t- t' |$ increases.

\bigskip


\noindent{\bf 4.3. Mean Integrated Squared Error}
\label{sec: MISE}

\smallskip

To measure the accuracy of an estimator $\hat{\mu}_N$ of $\mu_N$ over $[0,T]$, we use the Integrated Squared Error
 \begin{equation*}
\mathrm{ISE} = \int_{0}^T  \left( \hat{\mu}_{N}(t) - \mu_N(t) \right)^2 dt  .
\end{equation*}
As seen in Sections 2-3, the HT estimator (\ref{HT strat time-varying}) is unbiased and, when the sample sizes $n_h(t)$ are constant over time, 
 its  variance function is the same under the full and partial replacement designs  (in particular, for fixed panels). 
Therefore the HT estimator has the same mean integrated squared error \[ \mathrm{MISE}  =  \int_{0}^T   E\left( \hat{\mu}_{N}(t) - \mu_N(t) \right)^2 dt  \] under both designs. 
On the other hand, in comparison to fixed panels, 
the full and partial replacement designs can reduce 
the MISE by using suitable time-varying sample sizes $n_h(t)$. 
Specifically, the variance of $\hat{\mu}_{ht}(\tau_r), \,1\le r \le m,$ is minimal 
when $n_h(\tau_r)$ is chosen according to 
the  Neyman allocation rule: $n_h(\tau_r) =c_r N_h \sqrt{\gamma_h(\tau_r,\tau_r)} $ 
with the constant $c_r$ such that $\sum_{h} n_h(\tau_r)=n(\tau_r)$
(see e.g., Fuller, 2009, p. 21 for more details). 
Note that in practice, $\gamma_h(\tau_r,\tau_r)$ is unknown and must be estimated from the data.

\pagebreak


\setcounter{chapter}{5}
\setcounter{equation}{0} 
\noindent {\bf 5. Asymptotic results for the ISE}
\label{sec: main results}

\smallskip

We now determine the variance of the ISE for the HT estimator (\ref{HT strat time-varying}) under the full and partial replacement designs. 
We first write
\begin{align*}
\mathrm{Var}\left(\mathrm{ISE}\right)  & =
 \iint_{[0,T]^2} \mathrm{Cov} \left( \left\{ \hat{\mu}_{ht}(t) - \mu_N(t) \right\}^2 , \left\{ \hat{\mu}_{ht}(t') - \mu_N(t') \right\}^2 \right) dt dt' \\
& =  \frac{1}{N^4} \iint_{[0,T]^2}  \sum_{i,j,k,l \in U}\frac{  \Delta_{ijkl}(t,t')}{\pi_i(t)\pi_j(t)\pi_k(t')\pi_l(t') } \,
 X_i(t)X_j(t)X_k(t')X_l(t')
 \,  dt dt' , 
\end{align*}
 where 
 \begin{equation*}
 \Delta_{ijkl}(t,t') 
  =  \mathrm{Cov} \big(  \{ I_i (t) -\pi_i(t) \} \{ I_j (t) -\pi_j(t) \} \,, \,\{ I_k (t') -\pi_k(t') \} \{ I_l (t') -\pi_l(t') \} \big) .
\end{equation*}
Based on the independence of samples across strata, 
it can be shown that
\begin{equation}\label{Var ISE}
\begin{split}
&\mathrm{Var}\left(\mathrm{ISE}\right)  
 = \frac{1}{N^4}  \sum_{h=1}^H
  \iint_{[0,T]^2} \sum_{i,j,k,l \in U_h}  \frac{\Delta_{ijkl}(t,t')  }{f_{h}^2(t) f_h^2(t')} X_i(t)X_j(t)X_k(t')X_l(t') dt dt' \\
  & + \frac{2}{N^4} \sum_{h\ne h' }  \iint_{[0,T]^2} 
\sum_{i,k\in U_h} \frac{\Delta_{ik}(t,t') }{f_{h}(t)f_h(t') } \, X_i(t)X_k(t') \sum_{j,l \in U_{h'}} \frac{\Delta_{jl}(t,t')}{ f_{h'}(t) f_{h'}(t')} \, X_j(t)X_l(t') dt 
\end{split}
\end{equation}
The variance (\ref{Var ISE}) can be computed exactly if the $n_h(t)$ are constant over time  
but  requires large-sample approximations otherwise. 


\bigskip

\noindent{\bf 5.1. Asymptotic framework}

\smallskip

We let the strata sizes $N_h$, sample sizes $n_h(t)$, replacement rates $\alpha_h$ 
and number of replacements $m$ depend on the population size $N$ 
and let $N\to\infty$. 
The parameters $m, n_h(t), N_h $ go to infinity with $N $ while 
the number of strata and the observation period $[0,T]$ stay fixed.  
We make the following assumptions.

\begin{enumerate}

\item[(A1)] The curves $X_k, \, k  \ge 1,$ are integrable and 
uniformly bounded on $[0,T]$. 

 \item[(A2)] $\displaystyle \int_0^{\tau_r} g(t)dt = \frac{r}{m+1}\, , \, 0\le r \le m+1$, 
 where $g$ is a continuous, positive, and bounded function on $(0,T)$. 

\item[(A3)]  
For all $h$, the sampling rate function $f_h$ 
converges uniformly on $[0,T]$ to a continuous, positive limit function as $N\to\infty $. 

\item[(A4)]  For all $h$, 
the covariance function $\gamma_h$ converges uniformly on $[0,T]^2$ 
to a continuous limit as $N\to\infty $.

\item[(A5)] 
 $m=o(\min_h(N_h))$ as $N\to\infty$.

\end{enumerate}
\noindent
The number $H$ of strata, although fixed, can be large. The condition $N_h \to\infty$ is not restrictive as, typically, small strata $U_h$ are fully observed and do not contribute to the estimation error.  
(A1) allows discontinuity jumps in the individual curves $X_k$. 
However, (A4) requires that the strata covariance functions can be uniformly approximated by continuous functions, 
which entails that at any time $t$, only a negligible fraction of the $X_k(t)$ have discontinuity jumps. 
This assumption is needed under full replacement  to approximate the covariance $\gamma_h(t,t')$ 
by the variance $\gamma_{h}(t,t)$ around the diagonal $\{t=t'\}$. 
(A2) ensures that the replacement times are regularly spaced.  
 (A3) requires positive sampling rate functions, which is necessary for the consistent estimation of $\mu_N$. 
Finally (A5) is needed under partial replacement to approximate certain transition probabilities.

\bigskip



\noindent{\bf 5.2. Intermediate results}

\smallskip


 Let  $\tilde{X}_k(t) = X_k(t) - \mu_{h}(t)$ for $k\in U_h$ and $1\le h\le H$.  
Simple algebra yields 
\begin{equation}\label{decomp D_ijkl simplified}
\begin{split}
 \sum_{i,j,k,l\in U_h}& \Delta_{ijkl}(t,t') \, X_i(t)X_j(t)X_k(t')X_l(t')  \\ 
& =   \sum_{i,j,k,l \in U_h}    E \left(  I_i(t)I_j(t)I_k(t')I_l(t') \right)   \tilde{X}_i(t)\tilde{X}_j(t)\tilde{X}_k(t')\tilde{X}_l(t')   \\
& \qquad - N_h^2 \, f_{h}(t) f_{h}(t') \left(1-f_{h}(t)\right) \left(1-f_{h}(t')\right) \gamma_{h}(t,t) \,  \gamma_{h}(t',t') \, .
\end{split}
\end{equation}
%

The sum in the right-hand side of  (\ref{decomp D_ijkl simplified}) 
can be developed using the properties of SRSWOR.  
 Let $a_N \sim b_N$ denote the asymptotic equivalence of two real sequences $(a_N)$ and $(b_N)$. 

\begin{proposition}\label{prop: further decompose EI_ijkl X_ijkl} 
Assume either the full or the partial replacement design 
and (A1).
Let $i^\ast,j^\ast,k^\ast, l^\ast $ be four distinct units in a given stratum $U_h$. 
Then
\begin{align*}
& \sum_{i,j,k,l\in U_h}E\left( I_i(t)I_j(t)I_k(t')I_l(t')\right)  \tilde{X}_i(t) \tilde{X}_j(t)\tilde{X}_k(t')\tilde{X}_l(t') \\
& \quad \qquad \sim \left(  C_1(t,t')\, \gamma_{h}(t,t) \gamma_{h}(t',t') + C_2(t,t')  \gamma_{h}^2(t,t')\right)   N_h^2 
\end{align*}
uniformly in $t,t'\in [0,T]$ as $N_h \to\infty$, where 
%
%
\begin{align*}
C_1(t,t') & =   E\left( I_{i^\ast}(t) I_{k^\ast}(t') \right) 
 -  E\left( I_{i^\ast}(t)I_{j^\ast}(t)I_{k^\ast}(t')\right)  -  E\left( I_{i^\ast}(t)I_{k^\ast}(t')I_{l^\ast}(t')\right) \\ 
& \qquad +  E\left( I_{i^\ast}(t) I_{j^\ast}(t) I_{k^\ast}(t') I_{l^\ast}(t') \right)\\
\noalign{\noindent and} 
C_2(t,t')& =   2\,  E\left( I_{i^\ast}(t)I_{i^\ast}(t') I_{k^\ast}(t)I_{k^\ast}(t')\right)   -  4\, E\left( I_{i^\ast}(t) I_{i^\ast}(t') I_{j^\ast}(t) I_{k^\ast}(t')\right) \\
& \qquad + 2 \, E\left( I_{i^\ast}(t) I_{j^\ast}(t) I_{k^\ast}(t') I_{l^\ast}(t') \right).
\end{align*}

\end{proposition}

Under the full replacement design, the functions $C_1$ and $C_2$  can  be expressed in terms of $f_h(t)$ and $f_h(t')$ and $\mathrm{Var(ISE)}$ can readily be computed. Under partial replacement, an additional result is required. 
 Let $k$ and $l$ be two distinct units in a stratum $U_h$. Applying Lemma \ref{lem: s inter D MC} to $D=\{k,l\}$ and 
using the Chapman-Kolmogorov equations and large-sample approximations, we obtain the following transition probabilities.
%

\begin{proposition}\label{prop: transitionprob2}
Assume the  partial replacement design 
and (A3)-(A5). For all   
$0\le t \le t' \le T$, 
it holds as $N \to\infty $  that 
\[
\left\{ 
\begin{array}{l}
\medskip
 P  \left( k,   l \in s_h(t') \big|  k,l \in s_h(t) \right)  \sim \big[ \,(1-f_{h}(t)) \lambda_h(t,t') + f_{h}(t') \, \big]^2  , \\
\medskip
P\left( k,l \in s_h(t') \big|  k \in s_h(t),\, l \notin s_h(t) \right) \\
\medskip
\quad  \sim   \big[ - f_{h}(t) \left( 1-f_{h}(t) \right) \lambda_h^2(t,t') 
 + f_{h}(t') \left( 1-2\, f_{h}(t)\right) \lambda_h(t)
+ f_{h}^2(t') \big]  , \\
P \left( k,l \in s_h(t') \big|  k , l \notin s_h(t) \right) \sim \big[ \left(1-f_{h}(t)\right)\lambda_h(t,t') -\left( 1- f_{h}(t')\right)    \big]^2  .
\end{array}
\right. 
\]
\end{proposition}

\bigskip


\noindent {\bf 5.4. Variance of the Integrated Squared Error}
\label{subsec: Stab VarISE results}

\smallskip

Based on the previous findings, we can now state the main results. 

\begin{theorem}\label{Var ISE full}
Consider the HT estimator  (\ref{HT strat time-varying}) based on the full  replacement design. 
Assume  (A1), (A2), (A3) and (A4). Then as $N\to\infty$, 
\[
\mathrm{Var}\left(\mathrm{ISE} \right)
\sim \frac{2}{m N^2} \int_{0}^T \left( \sum_{h=1}^H \frac{N_h}{N}\, \frac{1-f_{h}(t)}{f_{h}(t)} \, \frac{1}{g(t)}\, \gamma_h(t,t) \right)^2  dt . 
\]

\end{theorem}

\smallskip


\begin{theorem}\label{Var ISE partial}
Consider the HT estimator  (\ref{HT strat time-varying}) based on the partial replacement design. 
Assume  (A1), (A2), (A3), and (A5). Then as $N\to\infty$, 
\[
\hspace*{-3.5mm}
\mathrm{Var}\left(\mathrm{ISE}  \right)
\sim \frac{2}{N^2} \iint_{[0,T]^2} \left( \sum_{h=1}^H \frac{N_h}{N} \,  \frac{1-f_{h}(t)}{f_{h}(t')} \,  \lambda_h(t,t') \,\gamma_h(t,t')  \right)^2 dtdt' . 
\]
\end{theorem}

Under additional assumptions, it is possible to find the asymptotic expression of $\lambda_h(t,t') $. 
Let $G$ be an antiderivative of the density $g$ in (A2).  
 
\begin{corol}\label{corol th4}
Assume the conditions of Theorem \ref{Var ISE partial} and suppose that 
(i) the  sample sizes $n_h(t)$ are constant over time, and
 (ii) $\lim_{N\to\infty} \left( \alpha_h m / (1-f_h)\right) = c_h <\infty$ exists. 
Then as $N\to\infty$,  
\[
\hspace*{-3.5mm}
\mathrm{Var}\left(\mathrm{ISE}  \right)
\sim \frac{2}{N^2} \iint_{[0,T]^2} \left( \sum_{h=1}^H \frac{N_h}{N}\, \frac{1-f_{h}}{f_{h}}\, \exp \left( - c_h \left| G(t)-G(t')\right| \right) \gamma_h(t,t') \right)^2 dtdt' . 
\]

\end{corol}
The previous condition (ii) is reasonable since $\alpha_h m/T$ is the average sample replacement rate per unit time, which in practice stays bounded. The symmetry of $\alpha_h$ and $m$ is conform to intuition, 
since multiplying either of these parameters by a given integer produces the same total of replaced units.  

Under the assumptions of Theorems \ref{Var ISE full}-\ref{Var ISE partial} and Corollary \ref{corol th4}, 
we now compare the  full and partial  replacement designs in terms of variability of the ISE. 
As in Section 4 we include fixed panels as a special case of partial replacement 
where $\alpha_h=c_h=0$. In comparison to fixed panels, 
partial replacement with $c_h>0$ induces an exponentially decreasing function in  $\mathrm{Var(ISE)}$. 
The decrease rate is larger when   $c_h$ is large and the data are highly positively correlated. 
In comparison to  partial replacements, 
the full replacement design divides the order of  $\mathrm{Var(ISE)}$ by a factor $m$, 
which massively stabilizes  the estimation performance.

\begin{remark}\label{estimation integral function}
If the survey's goals include evaluating  $I_N =  \int_0^T  \mu_N(t) dt $, 
then $ \hat{I}_{N} =  \int_0^T \hat{\mu}_{ht}(t) dt $ provides an unbiased estimator 
whose variance can be deduced from the previous results.  
As above, in comparison to fixed panels, 
 partial replacement of the sample reduces 
$\mathrm{Var}(\hat{I}_{N})$ by an exponentially decreasing function   
and full replacement divides the order of $\mathrm{Var}(\hat{I}_{N})$ by a factor $m$.  
\end{remark}


\bigskip

\setcounter{chapter}{6}
\setcounter{equation}{0} 
\noindent {\bf 6. Composite estimation}

\smallskip

The HT estimator (\ref{HT strat time-varying}) of $\mu_N(t)$ is only 
based on  current observations $X_k,\break  k\in s(t)$. 
The estimation can likely be improved by using past data in addition to current ones. 
Following the principle of composite estimation (e.g., Eckler, 1955), 
 we  utilize  the partial replacement design of Section 3 and recursively 
 define a new estimator $\hat{\mu}_c(t)$ as a linear combination of  $\hat{\mu}_{ht}(t)$ 
 and of 
  $\hat{\mu}_c(t-\delta) $ plus the estimated change in $\mu_N$ between $t-\delta$ and $t$, 
  where $\delta>0$ is a lag parameter to be specified.  
%

Let $0\le  t \le  t'  \le T$. If $|\nu(t)-\nu(t')|\le 1$,    the estimator 
\begin{align}\label{change estimator 1}
\widehat{\Delta}{\mu_N}(t,t') & = \frac{1}{N} \sum_{k\in U} \frac{I_k(t) I_k(t' )}{\pi_{kk}( t ,t'  )}  \left( X_k(t') - X_k(t) \right) 
\end{align}
of the level change  $\Delta\mu_N(t,t') = \mu_N(t') -\mu_N(t)$ is unbiased. 
If $| \nu(t) - \nu(t') | \ge 2 $, the previous estimator is extended as 
\begin{equation}
\widehat{\Delta}{\mu_N}(t,t') =
\widehat{\Delta}\mu_N(t , \tau_{\nu(t)+1}) + \sum_{r=\nu(t)+2}^{\nu(t')} \widehat{\Delta}\mu_N(\tau_{r-1},\tau_{r}) + \widehat{\Delta}\mu_N(\tau_{\nu(t')}, t'). 
\end{equation}

The composite estimator is defined by 
\begin{equation}\label{composite estimator}
\hat{\mu}_c(t) = 
\begin{cases}
\hat{\mu}_{ht}(t),  & 0 \le t < \tau_1, \\
Q \,\hat{\mu}_{ht}(t) + \left( 1-Q\right) \big( \hat{\mu}_c(t- \delta) + \widehat{\Delta}{\mu_N}(  t-\delta , t) \big)  , &
 \tau_1\le t \le T,
\end{cases}
\end{equation}
where $Q\in [0,1]$ must be specified and, by convention,   
$ \hat{\mu}_c (t) =  \hat{\mu}_c (0 )$ 
and $\widehat{\Delta}{\mu_N}(  t , t')=\widehat{\Delta}{\mu_N}(  0 , t')$ 
if $t < 0 \le t' $. 
Note that if $\alpha_h=0$ with  sample sizes $n_h(t)$ constant over time, 
if $\delta=0$, or if $Q=1$, then $\hat{\mu}_c(t)$ reduces to $\hat{\mu}_{ht}(t)$. 
The composite estimator is thus a shrinkage estimator 
whose parameters $\alpha_h$, $\delta,$ and $Q$ determine the relative importance of past and present data in the estimation.  

\bigskip



\setcounter{chapter}{7}
\setcounter{equation}{0} 
\noindent {\bf 7. Numerical study}

\smallskip

\begin{figure}[htbp]
\hspace*{-1cm}
\includegraphics[scale=.7]{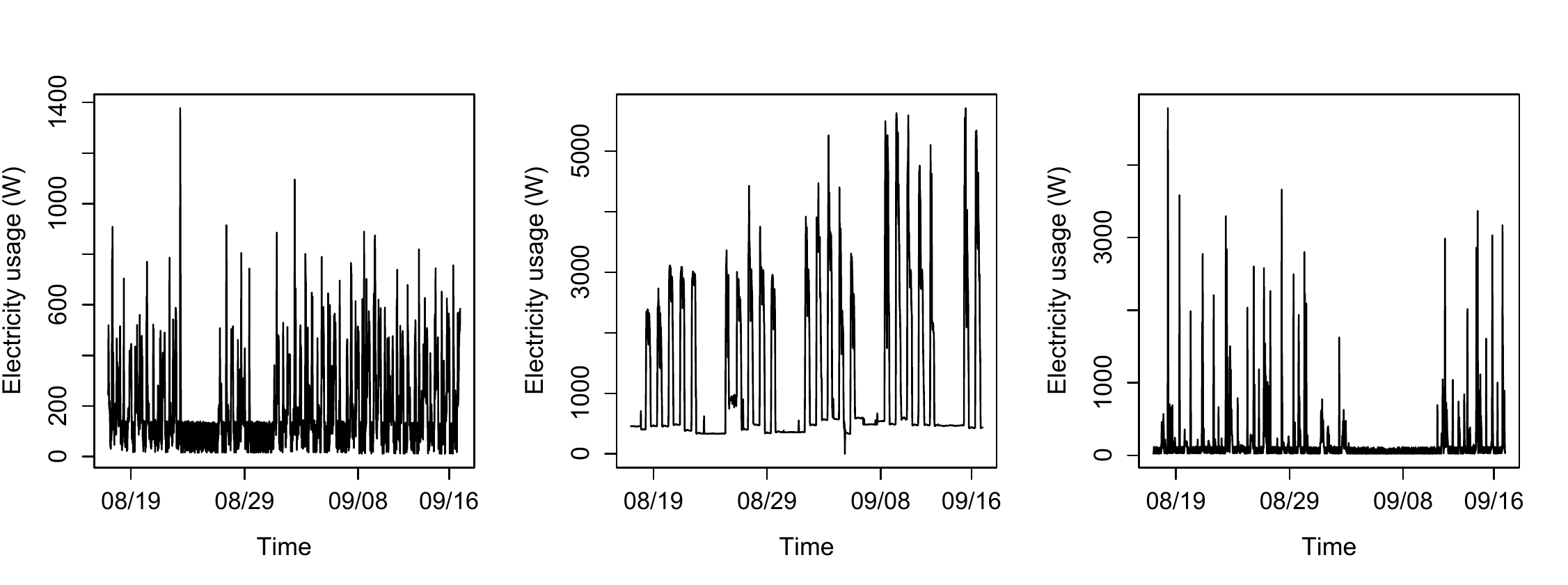}
\caption{Sample electricity curves.}
\label{sample curves}
\vspace*{-5mm}
\end{figure}

\begin{figure}[htbp]
\begin{center}
\includegraphics[scale=.54]{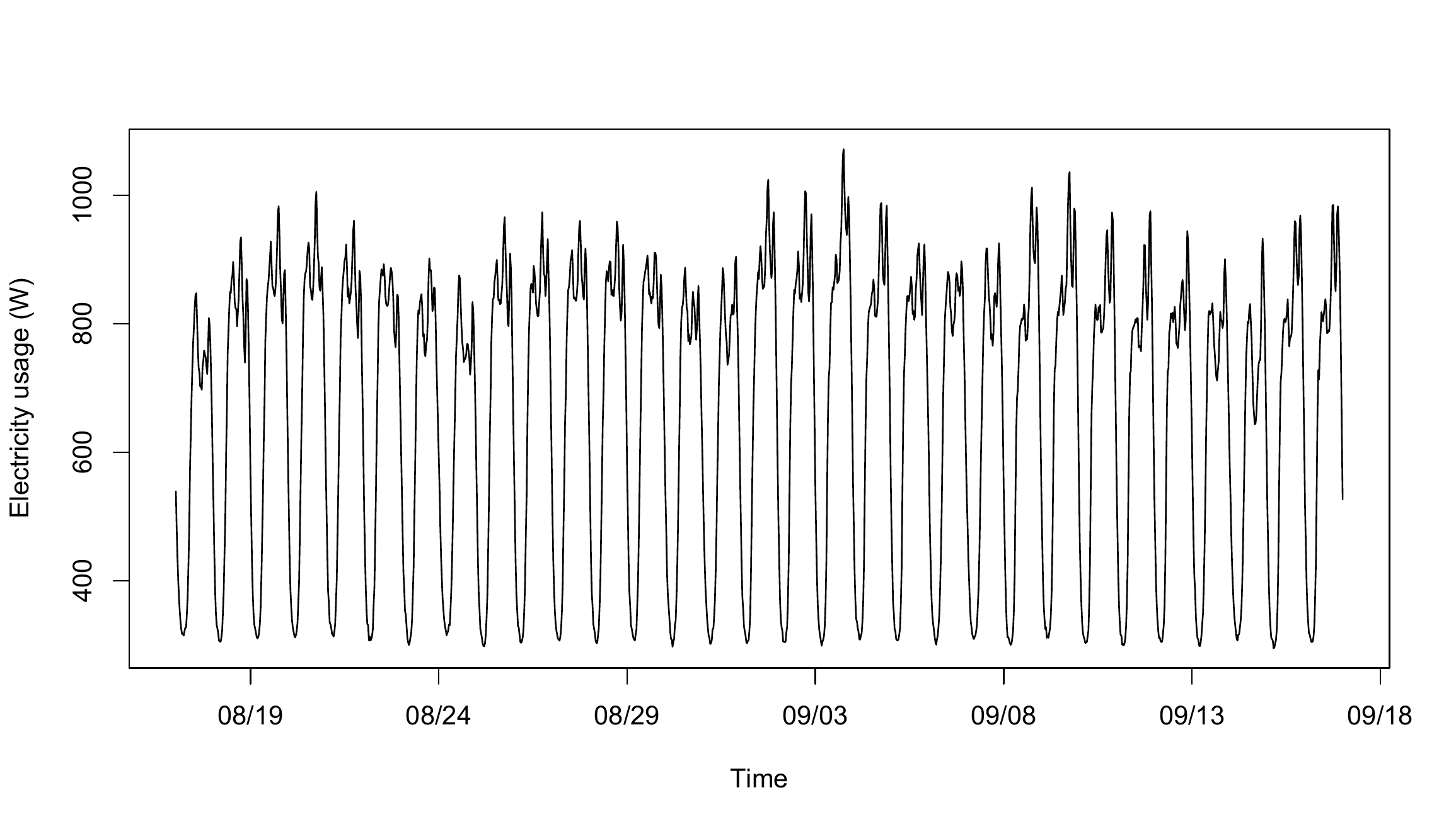}
\caption{Mean electricity consumption in the population.}
\label{mean curve}
\end{center}
\end{figure}

Here we examine the numerical performances of the HT estimator (\ref{HT strat time-varying}) and composite estimator (\ref{composite estimator}) based on the sampling designs of Section 3. We use electricity consumption data from the Irish CER Smart Metering Project conducted in 2009-10 (CER, 2011). During the project, smart meter readings (in kW) were collected every 30mn for $N=6445$ residential and business customers. (The data are available by request at  
www.ucd.ie/issda/data/commissionforenergyregulation/.)  
We focus on one month of data (8/17/2009-9/17/2009) and set the sampling rate to 5\% ($n = 322  $) for the whole period. 
Customer electricity curves and the population mean curve are displayed in Figures \ref{sample curves} and \ref{mean curve}. 
We stratify the population according to the type of contract (see Table \ref{strata sizes}) 
and replace the sample every 12 hours so that 
$\tau_r = 12 r, \, 1\le r \le m$ (in hours) with $m=61$ replacements. 

\begin{table}[htdp]
\vspace*{2mm}
\begin{center}
\begin{tabular}{c c c c}
\hline
Stratum & Residential & SME & Other \\ 
Size & 4225 & 485 & 1735 \\
\hline
\end{tabular}
\end{center}
\caption{Population strata. SME denotes Small-to-Medium Enterprises.}
\label{strata sizes}
\end{table}%

\begin{figure}[h]
\hspace*{-.7cm}
\includegraphics[width=1.1\linewidth]{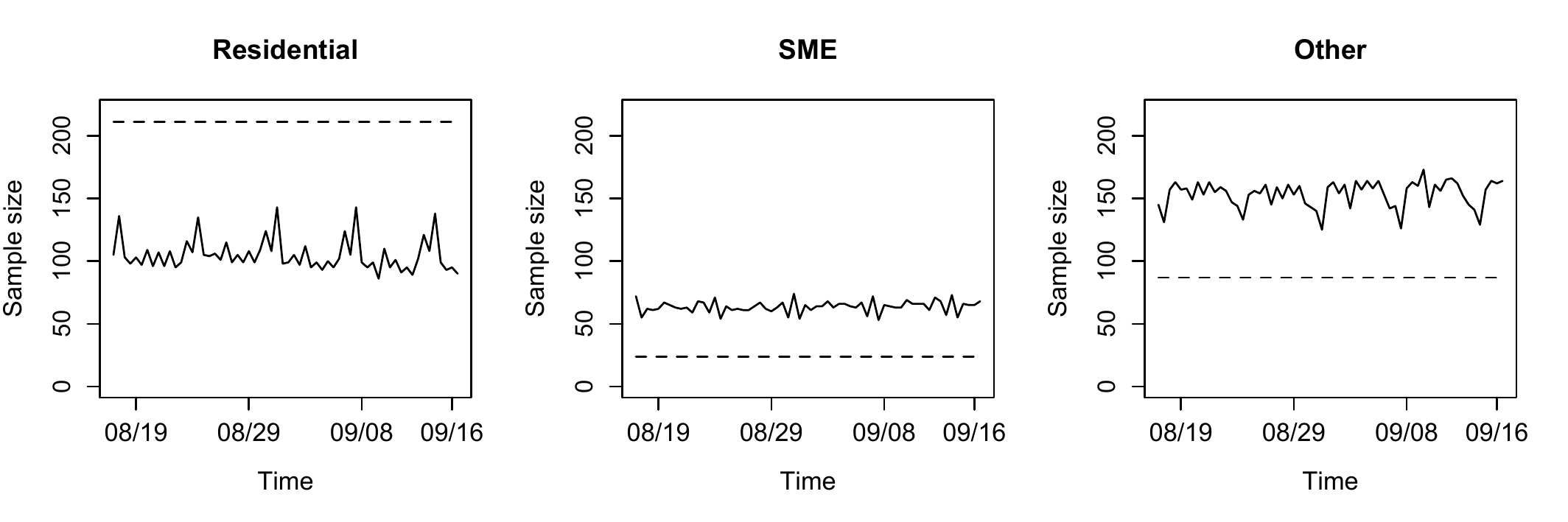}
\caption{Proportional (dashes)  and optimal  (solid line) sample allocation.}
\label{sample allocation}
\end{figure}

The study investigates three factors in the estimation: rotation design (full, partial or conventional), sample allocation (proportional, optimal or adaptive), and estimator (HT or composite).  
The conventional rotation design   
consists in specifying a rotation pattern (i.e. which population labels are in the sample at each time $\tau_r, 0\le r \le m$) and then randomly permuting the population labels (see e.g., Rao and Graham, 1964). 
The partial replacement- and conventional rotation designs adopt the same replacement rate $ \alpha\in\{ 0,0.1,0.2,\ldots,1 \}$ in each stratum $U_h, \,1\le  h \le 3$.   
In proportional allocation the sample sizes are $n_h = (N_h/N) n$ rounded to the nearest integer.   
Optimal (Neyman) allocation uses sample sizes $n_h(\tau_r) =c_r N_h \sqrt{\gamma_h(\tau_r,\tau_r)} $ 
with the constant $c_r$ such that $\sum_{h} n_h(\tau_r)=n$ (see Section 4.3). 
Figure \ref{sample allocation} illustrates the difference between these two allocations. 
Since the strata variances $\gamma_h(\tau_r,\tau_r)$ are unknown in practice, 
optimal allocation is infeasible; we use it as a benchmark. 
Adaptive allocation replaces the strata variances by estimates $\hat{\gamma}_h(\tau_r,\tau_r)$ in the optimal allocation method. 
We define $\hat{\gamma}_h(\tau_r,\tau_r)$ as the sample variance of the $X_k(t),\, k\in s_h(t),$ at the last observation time $t$ before $\tau_r$. 
In composite estimation we use the parameter values $\alpha \in\{ 0.1,0.2,\ldots,1 \}$, $\delta \in \{ 0.5, 1, 6, 12, 24\}$ (in hours), and $Q\in \{ 0,0.1,0.2,\ldots, 1\}$.  
For each combination of factors (sampling design, sample allocation, and estimator) and parameter values, we generate 
the time-varying sample $s = \{ s(t): t\in [0,T]\}$ by Monte Carlo simulation 
and compute the corresponding estimator and integrated squared error (ISE) 10,000 times.

\begin{figure}[h]
\begin{center}
\includegraphics[width=\linewidth]{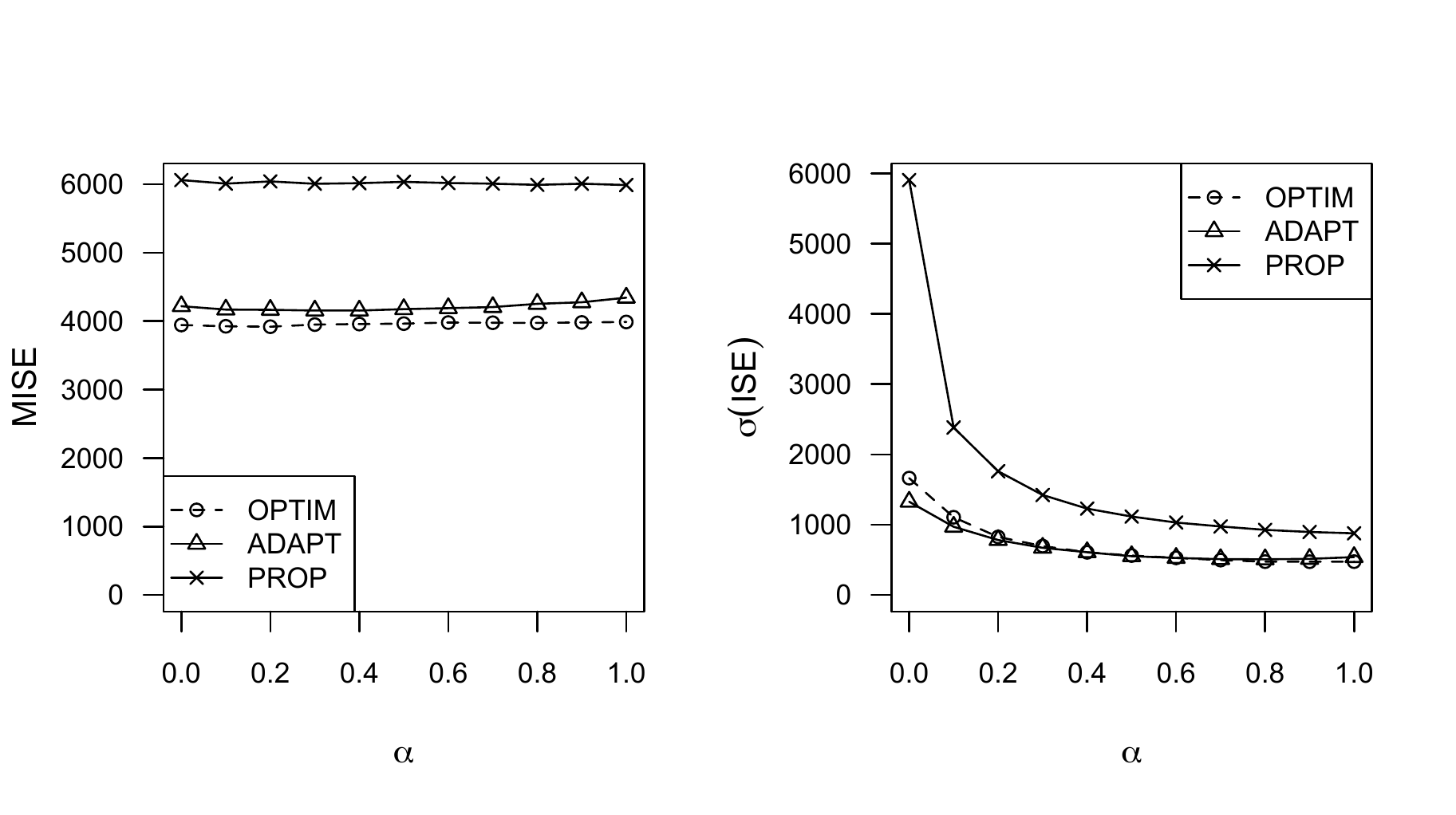}
\vspace*{-10mm}
\caption{Comparison of proportional (PROP), adaptive (ADAPT) and optimal (OPTIM) sample allocation 
using the Horvitz-Thompson estimator. 
The mean and standard deviation of the ISE are displayed in terms of the replacement rate $\alpha$}
\label{ISE graphs alloc}
\end{center}
\end{figure}

The mean and standard deviation of the ISE 
for the HT estimator are shown in Figure \ref{ISE graphs alloc}.    
This figure compares the three types of sample allocation under the partial replacement design. 
Numerical results for the full replacement design are nearly identical to partial replacement with $\alpha=1$. 
In line with Section 4.3 and Theorem 4, for a given  allocation, the mean integrated squared error (MISE) is the same for all $\alpha$ while the standard deviation $\sigma(\mathrm{ISE})$ is a decreasing function of $\alpha$. 
Unsurprisingly, proportional allocation gives far less accurate results than optimal allocation. 
Adaptive allocation is comparable to optimal allocation in terms of MISE, 
with a relative efficiency between 91\% and 95\% across the range of $\alpha$. 
Regarding $\sigma(\mathrm{ISE})$, adaptive allocation is superior to optimal allocation for $\alpha\le 0.6$. 
This result is not contradictory given that the optimal (Neyman) allocation is only optimal for the MISE and for fixed sample  sizes  
(note that adaptive allocation requires random sample sizes).

\begin{figure}[h]

\begin{center}
\includegraphics[width=\linewidth]{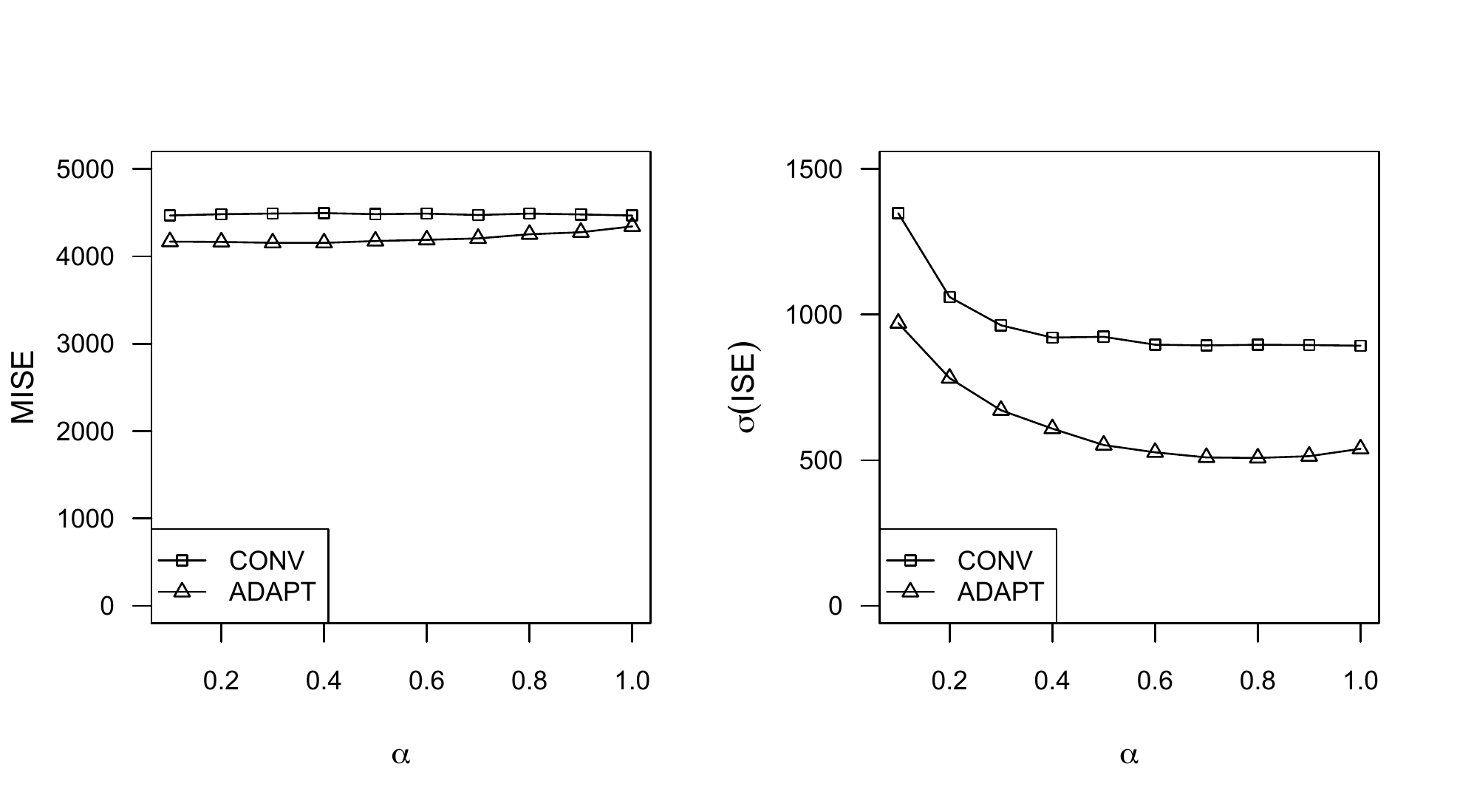}
\vspace*{-10mm}
\caption{Comparison of conventional rotation sampling (CONV) to our adaptive rotation designs (ADAPT)
using the Horvitz-Thompson estimator. The mean value and standard deviation of the ISE are displayed in terms of the replacement rate $\alpha$.}
\label{HT conventional vs MC}
\end{center}
\end{figure}

We now compare conventional rotation sampling to our rotation designs, using again the HT estimator. 
Under the conventional design, we employ proportional allocation before $\tau_1$ 
and take $n_h \propto N_h ( \int_0^{\tau_1} \hat{\gamma}_h(t,t)dt)^{1/2} $ afterwards, 
with $\sum_h n_h= n$ and  $\hat{\gamma}_h(t,t)$ being the sample variance of the $X_k(t),\, k\in s_h(t)$. 
The sample sizes $n_h$ approximately minimize the MISE over $[0,\tau_1]$ and yield a reasonable estimator 
$\hat{\mu}_{ht}(t)$ for $t\ge \tau_1$ provided that the strata variances $\gamma_h(t,t)$ do not vary excessively with respect to each other. 
For our rotation designs, we use the adaptive sample allocation described earlier in the section. 
As Figure \ref{HT conventional vs MC} shows, our rotation designs improve upon conventional rotation sampling 
by 3\% to 8\% for the MISE  and by 25\% to 45\% for $\sigma(\mathrm{ISE})$ across the range of  $\alpha$. 
We have also tried using the customers' monthly consumption  
to improve sample allocation under the conventional design. 
(This auxiliary information is readily available since customers are billed monthly.)
However, it did not increase the performances of conventional rotation.

\begin{figure}[htbp]
\begin{center}
\includegraphics[width=0.97 \linewidth]{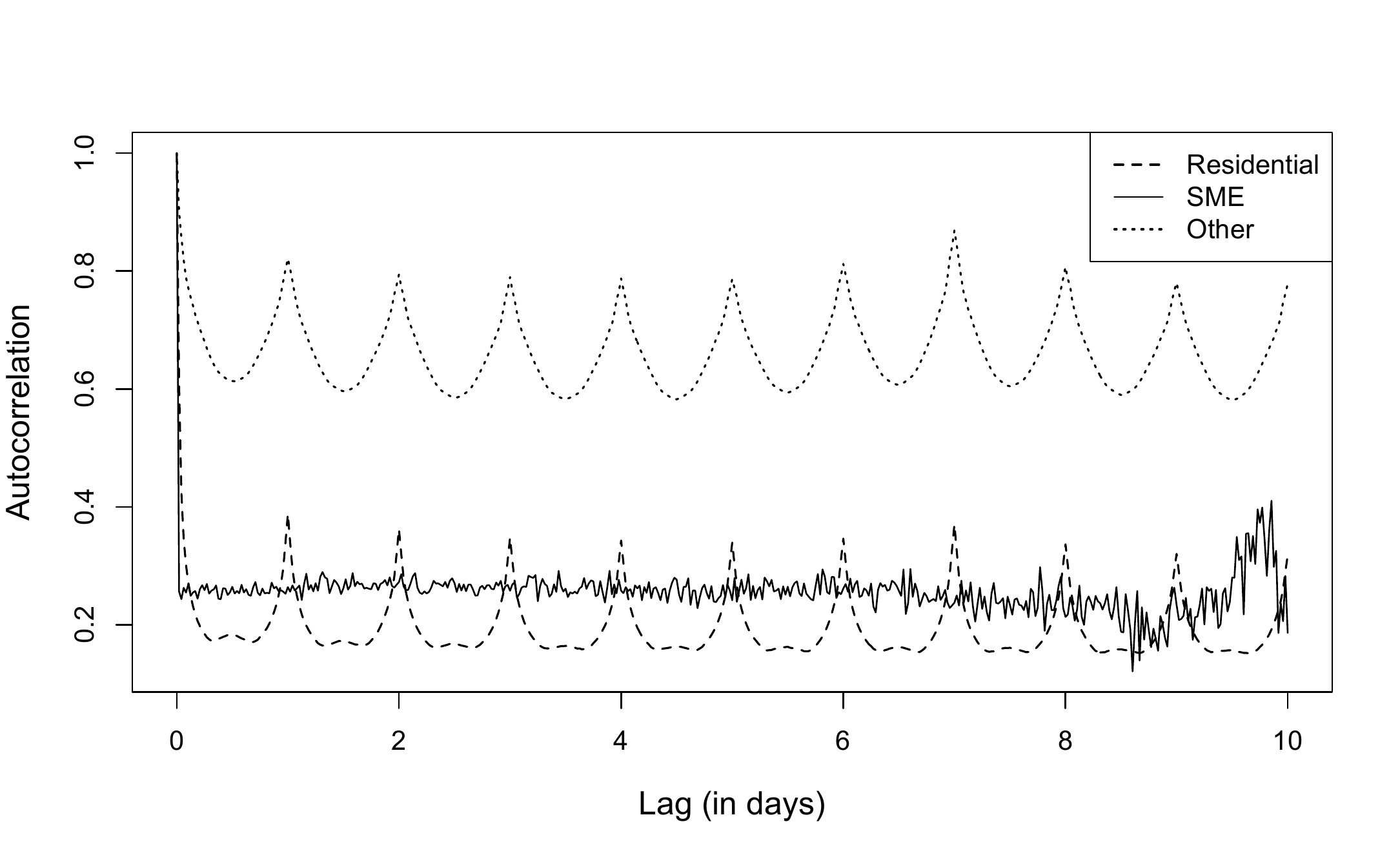}
\caption{Temporal dependence in electricity usage. The displayed functions  are the average autocorrelations 
$\Delta \mapsto \int_0^{T-\Delta} \rho_h(t,t+\Delta) dt$, where $\Delta$ is the time lag and  
 $\rho_h$ is the correlation function of stratum $U_h$. }
\label{fig: autocorrelation}
\end{center}
\end{figure}

Table \ref{results composite estimation} presents 
 results for the composite estimator $\hat{\mu}_c(t)$, 
which is defined in reference to a previous value $\hat{\mu}_c(t-\delta)$. 
As $\delta$ increases, 
$\alpha$ should decrease and $Q$ should increase in order to obtain the optimal MISE. 
In other words, if $\hat{\mu}_c(t)$ is defined with respect to a distant past $t-\delta$, then the sample should be more longitudinal 
(i.e., closer to a fixed panel) so that the change $\Delta\mu_N(t-\delta,t)$ is estimated more reliably. 
A large weight $Q$ should also placed on the current estimator $\hat{\mu}_{ht}(t)$ since for large $\delta$, the estimation of $\Delta\mu_N(t-\delta,t)$ is not very accurate. 
In comparison to the HT estimator (see Figure \ref{ISE graphs alloc}), the composite estimator brings no significant improvement (3\% at most for the MISE and 4\% at most for $\sigma(\mathrm{ISE})$ at any $\alpha$ value) although it uses more data. 
This is due to the overall weakness of the temporal dependence in electricity usage (see Figure \ref{fig: autocorrelation}). 
Since past data provide little information about current consumption,  
the composite estimator cannot greatly improve upon the HT estimator.  
In another numerical study with more strongly correlated data (not reported here),  
the composite estimator clearly dominated the HT estimator.

 \begin{table}[htdp]
\begin{center}
\begin{tabular}{c c c c}
$\delta$  & $\alpha_{opt}$ & $Q_{opt}$ & $\min(\mathrm{MISE})$ \\ 
\hline 
0.5h & 0.5 & 0.3 & 3872\\
1h & 0.5 & 0.4 & 3859\\
6h & 0.5 & 0.8 & 3850\\
12h & 0.4 & 0.8 & 3848 \\ 
24h & 0.2 & 1 & 3925\\
\hline
\end{tabular}
\end{center}
\caption{Composite estimation: optimal MISE and parameters $\alpha, Q$ in terms of  $\delta$.}
\label{results composite estimation}
\end{table}%


The results of this study have been seen to hold qualitatively for a range of  numbers  of replacements $m$. 
The full replacement design produces excellent estimates of $\int_0^T \mu_N(t)dt$ (see Remark \ref{estimation integral function}) with a relative error of 0.5\%.

\medskip

\setcounter{chapter}{8}
\setcounter{equation}{0} 
\noindent {\bf 8. Discussion}

\smallskip

In this paper we have devised rotation sampling designs for functional data. 
These survey designs are well suited to sensor network applications such as monitoring energy usage, internet traffic and TV/radio audiences. Unlike conventional rotation designs that specify the sample before the survey, our methodology allows for adaptive sampling. As theoretical and numerical results indicate, our approach produces better survey estimates than fixed panels and conventional rotation samples. The proposed composite estimator enhances the Horvitz-Thompson estimator by integrating both past and current data. Although both estimators yield comparable results in the numerical study, the composite estimator will likely be superior in the presence of stronger data correlation. 

The present work can be extended in several directions. First, our rotation designs can be modified to accommodate for instance cluster-, multistage- or PPS sampling, which could improve upon stratified sampling and SRSWOR. Second, in addition to adaptive sample allocation, other adaptive rules could increase the estimation accuracy. For example, in order to maximize the information content of the sample, the replacement rates could be based on the balance between longitudinal and cross-sectional variations in recent measurements. Third, a theoretical study of the composite estimator would facilitate statistical inference, help select the parameters $\delta$ and $Q$, and enable comparisons with the HT estimator. Finally, incorporating auxiliary information would profitably expand our approach. To this end, a comparison of design-based and model-assisted methods would be required to determine the most efficient integration scheme.


\medskip

\noindent {\bf Acknowledgment}
\smallskip

This research was supported in part by the Statistical and Applied Mathematical Sciences Institute (SAMSI). 
The author also thanks Herv\'e Cardot (Universit\'e de Bourgogne) 
and the referees for their helpful suggestions. 


\pagebreak



\appendix

\noindent {\bf Appendix}
\bigskip

\noindent {\bf Proof of Proposition \ref{prop: further decompose EI_ijkl X_ijkl}}

\smallskip

The sum under study can be decomposed as  $\sum_{\ell=1}^4 A_\ell (t,t')$,  
where 
\[
A_\ell(t,t') = \sum_{ \substack{i,j,k,l \in U_h \\  \mathcal{C}_{ijkl}=\ell}} E\left( I_i(t)I_j(t)I_k(t')I_l(t')\right)  \tilde{X}_i(t) \tilde{X}_j(t)\tilde{X}_k(t')\tilde{X}_l(t')  
\]
 and $\mathcal{C}_{ijkl}= \# \{ i,j,k,l\}$. 
To compute the $A_\ell$, 
we derive  $E\left( I_i(t)I_j(t)I_k(t')I_l(t')\right)$ 
based on the properties of SRSWOR 
and develop sums   $\sum  \tilde{X}_i(t) \tilde{X}_j(t)\tilde{X}_k(t')\tilde{X}_l(t')$ 
using the identity $\sum_{k\in U_h}\tilde{X}_k(t)=0$.
Let $i^\ast,j^\ast,k^\ast, l^\ast $ be four distinct units in $U_h$. 
To lighten the notation, we omit the subscript $k\in U_h$ in the sums to follow. 

We begin with the straightforward calculation of $A_1(t,t')$: 
\begin{equation}\label{prop2:A1}
A_1(t,t') =  E\big( I_{i^\ast}(t) I_{i^\ast}(t') \big) \sum \tilde{X}_k^2(t) \tilde{X}_k^2(t') . 
\end{equation}

The term $A_2(t,t')$ can be expressed as 
\begin{equation}
\begin{split}
&A_2(t,t')  =    E\big( I_{i^\ast}(t) I_{k^\ast}(t') \big)  
\Bigg[ (N_h-1)^2 \gamma_{h}(t,t)  \gamma_{h}(t',t') - \sum \tilde{X}_k^2(t) \tilde{X}_k^2(t') \Bigg] \\
&\ \ \  + 2\,  E\big( I_{i^\ast}(t)I_{i^\ast}(t') I_{k^\ast}(t)I_{k^\ast}(t')\big)   
\Bigg[   (N_h-1)^2 \gamma_{h}^2(t,t')  -   \sum \tilde{X}_k^2(t) \tilde{X}_k^2(t') \Bigg]   \\
& \ \ \ - 2\, \Big[ \, E\big( I_{i^\ast}(t) I_{i^\ast}(t')I_{k^\ast}(t')\big) + E\big( I_{i^\ast}(t)I_{k^\ast}(t) I_{k^\ast}(t')\big) \Big]
  \sum \tilde{X}_k^2(t) \tilde{X}_k^2(t')  .
  \end{split}
\end{equation}

Next, it can be shown that
\begin{equation}\label{prop2:A3}
\begin{split}
A_3(t,t')&  = \Big[\,   E\big( I_{i^\ast}(t)I_{j^\ast}(t)I_{k^\ast}(t')\big)  +  E\big( I_{i^\ast}(t)I_{k^\ast}(t')I_{l^\ast}(t')\big)  \Big] \\ 
& \qquad \times 
\Bigg[ - \left( N_h-1\right)^2 \gamma_{h}(t,t) \gamma_{h}(t',t') + 2 \sum  \tilde{X}_k^2(t) \tilde{X}_k^2(t') \Bigg] \\ 
&
\quad + 4\, E\big( I_{i^\ast}(t) I_{i^\ast}(t') I_{j^\ast}(t) I_{k^\ast}(t')\big)\\
& \qquad \times \Bigg[ - \left( N_h-1\right)^2 \gamma_{h}^2(t,t') + 2 \sum  \tilde{X}_k^2(t) \tilde{X}_k^2(t') \Bigg] .
\end{split}
\end{equation}

To compute  $A_4(t,t')$, 
 use the decomposition
\[
\sum_{i,j,k,l \in U_h} \tilde{X}_i(t) \tilde{X}_j(t)\tilde{X}_k(t')\tilde{X}_l(t') 
 = \sum_{\ell = 1}^4 \sum_{ \substack{i,j,k,l\in U_h \\ \mathcal{C}_{ijkl}=\ell }}  \tilde{X}_i(t) \tilde{X}_j(t)\tilde{X}_k(t')\tilde{X}_l(t')
\]
 together with (\ref{prop2:A1})-(\ref{prop2:A3})
 to obtain  
\begin{equation}\label{prop2:A4}
\begin{split}
&A_4(t,t')   =   E\big( I_{i^\ast}(t) I_{j^\ast}(t) I_{k^\ast}(t') I_{l^\ast}(t') \big) \times \\
&    \Bigg[ \left( N_h-1\right)^2 \gamma_{h}(t,t) \gamma_{h}(t',t')
 +2   \left(N_h-1\right)^2 \gamma_{h}^2(t,t')   
  - 6   \sum  \tilde{X}_k^2(t) \tilde{X}_k^2(t')\Bigg] .
\end{split}
\end{equation}

\medskip

The proof is completed by gathering (\ref{prop2:A1})--(\ref{prop2:A4}) 
and observing that all terms involving $ \sum  \tilde{X}_k^2(t) \tilde{X}_k^2(t')$ 
are of negligible order $\mathcal{O}(N_h)$ thanks to (A1). $\quad  \square$

\bigskip


\noindent{\bf Proof of Proposition \ref{prop: transitionprob2}}
\smallskip

Let $k,l $ be two distinct units in a stratum $U_h$. 
Consider the Markov chain 
$ \{(I_k + I_l)(\tau_r), 
\, r=0,\ldots,m\} $ which counts how many units among $k,l$ 
are present in the sample at the successive replacement times. 
This chain has three possible states: 0, 1, and 2.   
For  $1\le r \le m,$ the transition probability matrix 
\[ \mathbf{P}_r = \left( P \left( (I_k + I_l)(\tau_r) = j-1 \big|  (I_k + I_l)(\tau_{r-1}) = i-1 \right)\right)_{1\le i,j\le 3} \]
can be represented as  
\begin{equation}
\mathbf{P}_r =\mathbf{P}_r^\ast + \mathbf{E}_r  ,
\end{equation}
where 
\[
\mathbf{P}_r^\ast = 
\begin{pmatrix}
\medskip
{\left( 1-{\beta}_{r}\right) }^{2} & 2\,\left( 1-{\beta}_{r}\right) \,{\beta}_{r} & {\beta}_{r}^{2}\cr 
\medskip
\alpha_h \left( 1-{\beta}_{r}\right) &\alpha_h {\beta}_{r} +\left(1- \alpha_h\right)  \left( 1-{\beta}_{r}\right) & \left( 1-\alpha_h\right) {\beta}_{r} \cr
\medskip
\alpha_h^2 & 2 \left( 1 -\alpha_h\right) \alpha_h& \left( 1-\alpha_h\right)^2
\end{pmatrix} 
\]
and $\beta_r =  P \left(  k \in s_h(\tau_r) |k \notin  s_h(\tau_{r-1})\right) = \left(f_h(\tau_r) - (1-\alpha_h) f_h(\tau_{r-1}) \right) / \left( 1 - f_h(\tau_{r-1})\right)$. 
Recall that $\alpha_h =  P \left(  k \notin s_h(\tau_r) |k \in  s_h(\tau_{r-1})\right)$. 
The matrix $ \mathbf{E}_r$, whose  cumbersome expression is not given here, 
 is asymptotically negligible in comparison to $\mathbf{P}_r^\ast$.
More precisely, (A5) guarantees that  
$\max_{r =1,\ldots, m}\left\| \mathbf{E}_r \right\| = \mathcal{O}\left(1/N_h\right) $
 as $N \to\infty$, where $\| \cdot \| $ denotes an arbitrary matrix norm. 
For simplicity we use the spectral norm $\|\mathbf{A}\| = \sup_{\mathbf{x} \ne 0} \left( \mathbf{x'A'Ax}/\mathbf{x'x}\right)^{1/2} $ henceforth.

The transition probability matrices $\mathbf{P}_r $ have unit spectral norm. 
Using the binomial formula, the triangle inequality, and the inequality  $\| \mathbf{AB}\| \le \| \mathbf{A}\|\cdot \| \mathbf{B}\|$ 
holding for all compatible matrices $\mathbf{A}$ and $\mathbf{B}$,  
 it follows that 
\begin{align}\label{prop 4 maj 1}
\bigg\|   \prod_{r=\nu(t)+1}^{\nu(t')}  \mathbf{P}_r  -  \prod_{r=\nu(t)+1}^{\nu(t')} \mathbf{P}_r^\ast   \bigg\| 
& \le \sum_{r=1}^{ \nu(t') - \nu(t) -1  } { \nu(t') - \nu(t) -1 \choose r }   \left( \max_{q=\nu(t)+1,\ldots,\nu(t')}   \left\| \mathbf{E}_q \right\| \right)^r  \nonumber \\
& \le    \left( 1+   \max_{1\le r \le m}   \left\| \mathbf{E}_r \right\| \right)^m - 1 \nonumber \\
& =\mathcal{O} \left(  m  \max_{1\le r \le m}   \left\| \mathbf{E}_r \right\| \right) 
\end{align}
uniformly in $0\le t \le t' \le T$. Combining the previous results and (A5) yields 
\begin{equation}\label{approx prod P}
\prod_{r=\nu(t)+1}^{\nu(t')} \mathbf{P}_r =  \left(1 + o(1) \right) \prod_{r=\nu(t)+1}^{\nu(t')} \mathbf{P}_r^\ast  \,.
\end{equation}


We now study the simpler product $ \prod_{r=\nu(t)+1}^{\nu(t')} \mathbf{P}_r^\ast $. 
Without loss of generality, set $\nu(t)=0$ and  
$\nu(t')=r$. 
Define 
$\mathbf{Q}_r =  \prod_{l=1}^r \mathbf{P}_l^\ast  $ and write 
$ \left[ \mathbf{A} \right]_{ij} $ for 
the $(i,j)$th coefficient of a matrix $\mathbf{A}$. 
It remains  to compute 
$ [\mathbf{Q}_r]_{13}$, $[\mathbf{Q}_r ]_{23} $, and $[\mathbf{Q}_r ]_{33}$.   
\smallskip

Since $\mathbf{Q}_{r}$ is a transition probability matrix 
and $\left[ \mathbf{P}_r^\ast \right]_{2j} = \left[ \mathbf{P}_r^\ast \right]_{1j}^{1/2} \left[ \mathbf{P}_r^\ast \right]_{3j}^{1/2}$ 
for $j=1,3$, it can be shown by induction that    
$\left[ \mathbf{Q}_{r} \right]_{11}^{1/2} + \left[ \mathbf{Q}_{r} \right]_{13}^{1/2} =1$ for $1\le r \le m$.
%
%
As a result,  
$\left[ \mathbf{Q}_{r} \right]_{13}^{1/2} = \left( 1 -\beta_{r} - \alpha_h \right) \left[ \mathbf{Q}_{r-1} \right]_{13}^{1/2} + \beta_r$ and  
\begin{equation}\label{qr13_rec}
 \left[ \mathbf{Q}_{r} \right]_{13}^{1/2} - f_h(\tau_r)   = 
\frac{1-\alpha_h -f_h(\tau_r)}{1- f_h(\tau_{r-1})}  \left(  \left[ \mathbf{Q}_{r-1} \right]_{13}^{1/2} - f_h(\tau_{r-1}) \right) .
\end{equation}
Noting that 
$\left[ \mathbf{Q}_{1}\right]_{13}^{1/2}- f_h(\tau_1)  = - f_h(\tau_0)  ( 1-\alpha_h -f_h(\tau_1))/(1-f_h(\tau_0) ) $ and
iterating (\ref{qr13_rec}), 
 we obtain the identity 
$\left[ \mathbf{Q}_{r} \right]_{13}^{1/2} =  f_h(\tau_r) - f_h(\tau_0) \, \lambda_h(\tau_0, \tau_r)$. 


Similarly as above, we show that  
$\left[ \mathbf{Q}_{r} \right]_{33}^{1/2} = \left(1-f_h(\tau_0) \right) \lambda_h(\tau_0, \tau_r) + f_h(\tau_r) $.
(The expressions of $\left[ \mathbf{Q}_{r} \right]_{13}$ and $\left[ \mathbf{Q}_{r} \right]_{33}$ 
 can be checked by induction.)

Finally we turn to $\left[ \mathbf{Q}_{r} \right]_{23}$. 
The total probability formula yields 
\begin{equation}\label{total proba formula}
\hspace*{-1cm}
\begin{split}
P  \left( k,   l \in s(\tau_r) \right) 
&=  
 P  \left( k,   l \in s(\tau_r) \big|  k,l \in s(0) \right)   P  \left(   k,l \in s(0) \right) \\
&\ \ +2\, P\left( k,l \in s(\tau_r) \big|  k \in s(0),\, l \notin s(0) \right) P\left(  k \in s(0),\, l \notin s(0) \right) \\
&\ \ +P \left( k,l \in_h s_h(t') \big|  k , l \notin s(0) \right)P \left(  k , l \notin s(0) \right).
\end{split}
\end{equation}

\pagebreak

Based on (\ref{approx prod P}) we obtain 
\begin{equation}\label{total proba asymp}
\hspace*{-9mm}f_h(\tau_r)^2 \sim \left( f_h(\tau_0)^2   \left[ \mathbf{Q}_{r} \right]_{11} + 2\, f_h(\tau_0) \left(1-f_h(\tau_0)\right)  \left[ \mathbf{Q}_{r} \right]_{21} + \left(1-f_h(\tau_0) \right)^2  \left[ \mathbf{Q}_{r} \right]_{31} 
\right).
\end{equation}

The proof  is completed by 
expressing $\left[ \mathbf{Q}_{r} \right]_{13}$ and $\left[ \mathbf{Q}_{r} \right]_{33}$
 in (\ref{total proba asymp}). \quad $\square$

\bigskip


\noindent{\bf Proof of Theorem \ref{Var ISE full}}
\smallskip

We start by finding the asymptotic expressions of $C_1(t,t')$ and $C_2(t,t')$ in Proposition \ref{prop: further decompose EI_ijkl X_ijkl}.  
In view of (A3), the properties of SRSWOR and the independence of $s_h(\tau_r),\, 1\le r \le m,$ 
under full replacement, it comes that 
\[
\left\{
\begin{array}{l}
C_1(t,t')  \sim    f_{h}(t) f_{h}(t') \left( 1-  f_{h}(t)\right) \left( 1-  f_{h}(t') \right) \\
C_2(t,t') \sim  2\,  f_{h}(t) f_{h}(t') \left( 1-  f_{h}(t)\right) \left( 1-  f_{h}(t') \right) \delta_{\nu(t) \nu(t')} 
\end{array}
\right.
\]
uniformly in  $t,t'\in [0,T]$ as $N\to\infty $. 
Hence, the last term in  (\ref{decomp D_ijkl simplified})
cancels out with the term in $C_1(t,t')$ of Proposition \ref{prop: further decompose EI_ijkl X_ijkl}.

Writing $ \mathrm{Var(ISE)} = (2/N^2) \iint_{[0,T]^2} \phi_N(t,t') dtdt'$,
we deduce that 
\begin{equation}\label{equiv phi_N}
\phi_N(t,t')  \sim  \left( \sum_{h=1}^H \frac{N_h}{N} \, \frac{1-f_{h}(t)}{f_h(t')}\, \delta_{\nu(t)\nu(t')} \gamma_{h}(t,t')  \right)^2 
\end{equation}
uniformly in $t,t' \in [0,T]$  as $N\to\infty$. Using (A2)-(A4), the mean value theorem, 
integral approximations and a change of variable, 
we obtain 
\begin{align*}
\mathrm{Var(ISE)} & \sim  \frac{2}{N^2} \sum_{h,h'} \frac{N_h N_{h'}}{N^2} \sum_{r=1}^{m+1}  \frac{\left(1-f_{h}(\tau_r)\right) \left(1-f_{h'}(\tau_r)\right)
}{f_h(\tau_r) f_{h'}(\tau_r)} \iint_{[\tau_{r-1},\tau_{r}]^2}  \gamma_{h}(t,t')  \gamma_{h'}(t,t') dtdt' \\ 
&  \sim  \frac{2}{N^2} \sum_{h,h'} \frac{N_h N_{h'}}{N^2} \sum_{r=1}^{m+1}  \frac{\left(1-f_{h}(\tau_r)\right) \left(1-f_{h'}(\tau_r)\right)
}{f_h(\tau_r) f_{h'}(\tau_r)}  \left(\tau_{r} - \tau_{r-1}\right)^2   \gamma_{h}(\tau_r,\tau_r)  \gamma_{h'}(\tau_r,\tau_r)  \\
&  \sim  \frac{2}{N^2} \sum_{h,h'} \frac{N_h N_{h'}}{N^2} \sum_{r=1}^{m+1}  \frac{\left(1-f_{h}(\tau_r)\right) \left(1-f_{h'}(\tau_r)\right)
}{f_h(\tau_r) f_{h'}(\tau_r)} \, \frac{1}{m^2 \,g(\tau_{r})^2}\,   \gamma_{h}(\tau_r,\tau_r)  \gamma_{h'}(\tau_r,\tau_r)  \\
&  \sim  \frac{2}{mN^2} \sum_{h,h'} \frac{N_h N_{h'}}{N^2} \int_0^T  \frac{\left(1-f_{h}(t)\right) \left(1-f_{h'}(t)\right)
}{f_h(t) f_{h'}(t)} \, \frac{1}{ g(t)} \,   \gamma_{h}(t,t)  \gamma_{h'}(t,t)  dt \, . \quad  \square 
\end{align*}

\pagebreak


\noindent{\bf Proof of Theorem \ref{Var ISE partial}}
\smallskip

This result is established along the same lines as Theorem \ref{Var ISE full}. 
 In view of
 Lemmas   \ref{lem: s inter D MC}-\ref{r-step transition probability} and Proposition \ref{prop: transitionprob2},  
it holds for all $0\le t \le t' \le T$ that as $N\to\infty $,
\[
\begin{split}
 E\left( I_{i^\ast}(t) I_{k^\ast}(t') \right) &
 = P\left( k^\ast \in s(t') \big|   i^\ast, k^\ast \in s(t) \right)   P\left(  i^\ast , k^\ast \in s(t) \right)  \\
&\qquad + P\left( k^\ast \in s(t') \big|   i^\ast \in s(t) , k^\ast \notin s(t) \right)   P\left(  i^\ast  \in s(t), \, k^\ast \notin s(t) \right)  \\
& = P\left( k^\ast \in s(t') \big|  k^\ast \in s(t) \right)   P\left(  i^\ast , k^\ast \in s(t) \right)  \\
&\qquad  + P\left( k^\ast \in s(t') \big|  k^\ast \notin s(t) \right)   P\left(  i^\ast  \in s(t), \, k^\ast \notin s(t) \right)  \\
& \sim \big[  \left( 1 - f_{h}(t) \right) \lambda_h(t,t') + f_{h}(t')\,  \big]  f_{h}^2(t)   \\
&\qquad +\big[ \,  f_{h}(t') - f_{h}(t)\lambda_h(t,t') \, \big] f_{h}(t) \left( 1-f_{h}(t) \right) \\
& = f_{h}(t) f_{h}(t') .
\end{split}
\]
 By symmetry this equality holds for all $t,t'\in [0,T]$.   
Similarly, we find that 
\[\label{asymp eq EI}
\left\{ 
\begin{array}{lll}
  E\left( I_{i^\ast}(t)I_{j^\ast}(t)I_{k^\ast}(t')\right)  & \sim &  f_{h}^2(t) f_{h}(t'), \\
    E\left( I_{i^\ast}(t)I_{k^\ast}(t')I_{l^\ast}(t')\right) & \sim &  f_{h}(t) f_{h}^2(t'), \\
  E\left( I_{i^\ast}(t) I_{j^\ast}(t) I_{k^\ast}(t') I_{l^\ast}(t') \right) & \sim &  f_{h}^2(t) f_{h}^2(t') ,\\
%
  E\left( I_{i^\ast}(t)I_{i^\ast}(t') I_{k^\ast}(t)I_{k^\ast}(t')\right)   
  & \sim &  \big[ \,(1-f_{h}(t)) \lambda_h(t,t') + f_{h}(t') \, \big]^2 f_{h}^2(t)  , \\
    E\left( I_{i^\ast}(t) I_{i^\ast}(t') I_{j^\ast}(t) I_{k^\ast}(t')\right) 
 & \sim &  \big[ \left( 1-f_{h}(t)\right) \lambda_h(t,t') + f_{h}(t' ) \, \big] \, f_{h}^2(t)  f_{h}(t') . \\ 
\end{array}
\right. 
\]
Therefore
\begin{equation*}
\left\{
\begin{array}{l}
C_1(t,t')  \sim    f_{h}(t) f_{h}(t') \left( 1-  f_{h}(t)\right) \left( 1-  f_{h}(t') \right)\\
C_2(t,t') \sim  2\, f_{h}^2(t)  \left( 1-f_{h}(t)\right)^2 \lambda_h^2(t,t') 
\end{array}
\right. \: .
\end{equation*}

 Combining this result and Proposition \ref{prop: further decompose EI_ijkl X_ijkl}, 
it stems from (\ref{decomp D_ijkl simplified}) that 
\begin{equation}\label{as lem2}
\begin{split}
 \sum_{i,j,k,l\in U_h} & \Delta_{ijkl}(t,t') \, X_i(t)X_j(t)X_k(t')X_l(t') \\
&\qquad \sim  2\, f_{h}^2(t)  \left( 1-f_{h}(t)\right)^2 \lambda_h^2(t,t') \gamma_h^2(t,t') N_h^2 \, . 
\end{split}
\end{equation}

Based on Theorem \ref{covariance estimator: partial},
the last term in 
(\ref{Var ISE}) computes as 
\begin{equation}\label{interstrata contrib}
\sum_{i,k\in U_h} \frac{\Delta_{ik}(t,t') }{f_{h}(t)f_h(t') } \, X_i(t)X_k(t') = 
  N_h\, \frac{1  - f_{h}(t )}{f_{h}(t')}\, \gamma_{h}(t,t') \, \lambda_h(t,t') .
\end{equation}

Writing $ \mathrm{Var(ISE)} = (2/N^2) \iint_{[0,T]^2} \phi_N(t,t') dtdt'$,
(\ref{as lem2})-(\ref{interstrata contrib}) imply that  
\begin{equation}\label{equiv phi_N2}
\phi_N(t,t') \sim    \Bigg( \sum_{h=1}^H \frac{N_h}{N} \, \frac{1-f_{h}(t)}{f_h(t')}\,  \gamma_{h}(t,t') \, \lambda_h(t,t') \Bigg)^2 
\end{equation}
for all $t,t'\in [0,T]$ as $N\to\infty$.
To apply the dominated convergence theorem, it suffices  to check that the $\phi_N, \, N\ge 1,$ 
are uniformly bounded on $[0,T]^2$. 
Now, the right-hand side of (\ref{equiv phi_N2}) has a finite number of terms. 
In view of (A3)-(A4),  
$(1-f_{h}(t))/f_h(t')$ and $\gamma_h(t,t') $ are uniformly bounded with respect to $t,t'\in [0,T]$ and $N$. 
Finally, $|\lambda_h(t,t')| \le 1$ as a product of eigenvalues of transition probability matrices, which  concludes the proof. $\quad \square$

\bigskip


\noindent{\bf Proof of Corollary \ref{corol th4}}
\smallskip

It suffices to show that 
$\lambda_h(t,t') \sim \exp \left( - c_h \left| G(t)-G(t')\right| \right)$ 
as $N\to\infty$.  
Condition (i) yields $\lambda_h(t,t') =   \left( 1 - \alpha_h/(1 - f_{h}) \right)^{|\nu(t)-\nu(t')|} $ 
and 
$|\nu(t)-\nu(t')|= (m+1) \left|G(\tau_{\nu(t)})- G(\tau_{\nu(t')}) \right| \sim  m \left|G(t)- G(t') \right|$ by (A2). 
Condition (ii) and the approximation $\ln(1 +x) \sim x $ as $x\to 0$ 
provide the conclusion. $\quad \square$


\bigskip

\noindent{\large\bf References}


\begin{description}
\item
Cardot, H., Chaouch, M., Goga, C., and Labru{\`e}re, C. (2010).
\newblock Properties of design-based functional principal components analysis.
\newblock {\em J. Statist. Plann. Inference}, {\bf 140}, 75-91.

\item 
Cardot, H., Degras, D., and Josserand, E. (2012).
\newblock Confidence bands for {H}orvitz-{T}hompson estimators using sampled
  noisy functional data.
\newblock {\em Bernoulli.}
\newblock Forthcoming, also available at http://arxiv.org/abs/1105.2135.

\item  Cardot, H., Camelia Goga, C., and Lardin, P. (2013). 
\newblock Uniform convergence and asymptotic confidence bands for model-assisted
estimators of the mean of sampled functional data. 
\newblock {\em Electron. J.  Statist.}, {\bf 7}, 562-596. 


\item Chiky, R., Cubill{\'e}, J., Dessertaine, A., H{\'e}brail, G., and Picard, M.-L.
  (2008).
\newblock {\'E}chantillonnage spatio-temporel de flux de donn{\'e}es
  distribu{\'e}s.
\newblock In Guillet, F. and Trousse, B., editors, {\em EGC'08}, pp.
  169--180.

\item Commission for Energy Regulation (2011). 
\newblock Smart metering information paper 4 (CER 11/080). 
\newblock  Dublin, Ireland.  http://www.cer.ie.

\item Degras, D. (2011).
\newblock Simultaneous confidence bands for nonparametric regression with
  functional data.
\newblock {\em Statist. Sinica}, {\bf 21}, 1735-1765.

\item Eckler, A.~R. (1955).
\newblock Rotation sampling.
\newblock {\em Ann. Math. Statist.}, {\bf 26}, 664-685.


\item Fuller, W.~A. (2009).
\newblock {\em {Sampling statistics.}}
\newblock {Wiley Ser. Surv. Methodol.. Hoboken, NJ: John Wiley \& Sons.}

\item Horvitz, D.~G. and Thompson, D.~J. (1952).
\newblock A generalization of sampling without replacement from a finite
  universe.
\newblock {\em J. Amer. Statist. Assoc.}, {\bf 47}, 663-685.

\item Lavall\'ee, P. (1995). 
\newblock Cross-sectional weighting of longitudinal surveys of individuals and households using the weight share method. 
\newblock {\em Surv. Methodol.}, {\bf 21}, 25-32.



\item 
Rao, J. N.~K. and Graham, J.~E. (1964).
\newblock Rotation designs for sampling on repeated occasions.
\newblock {\em J. Amer. Statist. Assoc.}, {\bf 59}, 492-509.


\item Thompson, S.~K.; Seber, G.~A. (1996). 
\newblock {\em Adaptive sampling}. 
\newblock 
John Wiley \& Sons, Inc., New York. 


\item Wolter, K.~M. (1979).
\newblock Composite estimation in finite populations.
\newblock {\em J. Amer. Statist. Assoc.}, {\bf 74}, 604-613.

\end{description}


%
%
%

\vskip .65cm
\noindent
DePaul University
\vskip 2pt
\noindent
E-mail: ddegrasv@depaul.edu

\vskip .3cm
\end{document}